\documentclass[12pt]{spieman}
\usepackage{amsmath,amsfonts,amssymb}
\usepackage{graphicx}
\usepackage{setspace}
\usepackage{tocloft}

\usepackage{libertinus,libertinust1math} 
\usepackage{booktabs} 
\usepackage{orcidlink} 
\usepackage[
    separate-uncertainty=true,
    exponent-product=\cdot,
    detect-weight=true,
    ]{siunitx} 
\DeclareSIUnit\photon{ph}
\DeclareSIUnit\parsec{pc}
\usepackage{multirow, makecell} 
\usepackage[version=4]{mhchem} 
\usepackage{physics}
\newcommand{\bigO}{\mathcal{O}} 
\newcommand{\mean}[1]{\langle {#1}\rangle}
\newcommand{\R}{\mathbb{R}}
\newcommand{\iprod}[2]{\left\langle #1 \! \mid \! #2 \right\rangle}
\usepackage{csquotes}
\usepackage[capitalise]{cleveref} 
\usepackage{caption}
\usepackage{subcaption}
\title{The Nulling Interferometry Cryogenic Experiment (NICE): Architecture, requirements, and preliminary warm precursor results}

\author[a*$\dagger$\orcidlink{0000-0002-2982-0390}]{Thomas Birbacher}
\author[a$\dagger$\orcidlink{0000-0003-3992-342X}]{Jonah T. Hansen}
\author[a\orcidlink{0000-0002-5476-2663}]{Felix A. Dannert}
\author[a\orcidlink{0000-0001-6282-1339}]{Germain Garreau}
\author[a\orcidlink{0000-0001-9250-1547}]{Adrian M. Glauser}
\author[a]{Ryan Meierhofer}
\author[a]{Julio Pino Jiménez}
\author[a\orcidlink{0000-0001-9498-7958}]{Mohanakrishna Ranganathan}
\author[a,b\orcidlink{0000-0003-3829-7412}]{Sascha P. Quanz}
\affil[a]{\small ETH Zürich, Institute for Particle Physics \& Astrophysics, Wolfgang-Pauli-Str. 27, 8093 Zurich, Switzerland}
\affil[b]{\small ETH Zürich, Department of Earth and Planetary Sciences, Sonneggstrasse 5, 8092 Zurich, Switzerland}

\cftpagenumbersoff{figure}
\cftpagenumbersoff{table}
\begin{document}
\maketitle

\begin{abstract}
The success of the Large Interferometer For Exoplanets (LIFE) space mission depends on measuring the faint mid-infrared emission spectra of exoplanets while suppressing the glare of a host star.
This requires an instrument capable of high-contrast nulling interferometry with exceptional sensitivity.
While previous testbeds have proven the principle of deep, stable nulls, they have not combined high contrast with the high throughput and cryogenic operation required for LIFE.
Here, we present the architecture of the Nulling Interferometry Cryogenic Experiment (NICE), a mid-infrared nulling testbed, to increase the technological readiness of LIFE.
We derive the laboratory requirements necessary to validate the LIFE beam combiner and present the optical design of NICE.
Finally, we report results from the ambient \enquote{Warm Bench} precursor, which has successfully demonstrated the required null depth ($<\num{e-5}$) using a polarized narrowband \SI{4.7}{\um} source, and the required throughput ($> \SI{17}{\percent}$) using one of the two nulling channels.
\end{abstract}


{\noindent \footnotesize\textbf{*}Send correspondence to Thomas Birbacher, \linkable{thomabir@phys.ethz.ch}.\newline
\textbf{$\dagger$} Joint first authors for this article.}

\begin{spacing}{1}   

\section{Introduction}

The Large Interferometer For Exoplanets\cite{Quanz_2022} (LIFE) space mission is currently the most promising mission concept for systematically characterizing the atmospheres of terrestrial exoplanets in the mid-infrared.
LIFE aims to measure the emission spectra of $\approx \num{50}$ temperate terrestrial exoplanets in the habitable zones of nearby solar-type stars, and thus quantify the presence of a large set of spectral biosignatures\cite{Konrad_2022}.

To overcome the contrast between planet and star, LIFE is designed as a nulling interferometer, a concept originally proposed by Bracewell\cite{Bracewell_1978} for exoplanet detection.
In its simplest form, it consists of two telescope apertures that combine their light such that the on-axis star is nulled out through destructive interference, while the off-axis planet undergoes constructive interference.
The reference design\cite{Glauser_2024} for LIFE is a Double-Bracewell nuller\cite{Angel_1997}, with four formation-flying collector telescopes and a central combiner spacecraft.

Because of the faintness of temperate terrestrial exoplanets ($\approx \SI[per-mode=repeated-symbol]{1}{\photon\per\s\per\m\squared\per\um}$ at \SI{10}{\um} for an Earth-twin at \SI{10}{\parsec})\cite{Dannert_2022},
and the targeted wavelength range in the mid-infrared (\SIrange{4}{18.5}{\um}), LIFE depends on a space-based, cryogenic, high-throughput instrument to achieve the required sensitivity.
To reach a null deep enough for the characterization of such a planet ($\approx \num{e-5}$ raw null depth before post-processing), optical path-length differences in the instrument have to be controlled to the nanometer-level, and intensity and polarization mismatches reduced to the sub-percent level.
Combined, these factors necessitate an exceptionally stable optical setup with high demands on throughput and alignment accuracy, housed in a low-vibration cryostat.

\begin{figure}
    \centering
    \includegraphics[]{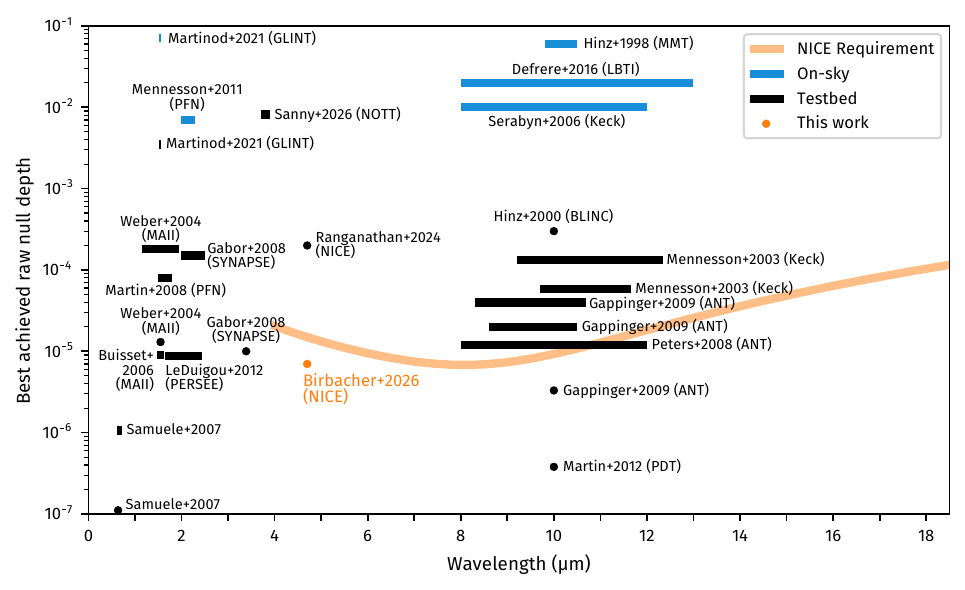}
    \caption[History of best measured nulls]{A sample of the best published raw null depths with on-sky nulling facilities and testbeds, compared with the requirement for NICE, derived in \cref{sec:nice1-requirements}. Rectangular markers denote nulls with more than \SI{2}{\percent} spectral bandwidth, and circles denote narrowband nulls.}
    \label{fig:nice1_history}
\end{figure}

Attempts towards such an instrument have been made before, notably for ESA's Darwin\cite{Kaltenegger_2005,Cockell_2009} and NASA's TPF-I\cite{Beichman_1999,Martin_2005} missions.
From ESA's side, the SYNAPSE\cite{Gabor_2008_Tests}, MAII\cite{Weber_2004,Buisset_2006}, and PERSEE\cite{LeDuigou_2017} testbeds have measured stable nulls deeper than \num{e-5} from \SIrange{1.1}{3.4}{\um}.
JPL's Planet Detection Testbed\cite{Martin_2012} (PDT) has achieved $< \num{e-6}$ nulls at \SI{10}{\um} in combination with a detection of a simulated planet, and the Achromatic Nulling Testbed\cite{Peters_2008,Gappinger_2009} (ANT) has demonstrated a \num{e-5} null covering a broad spectral band from \SIrange{8}{12}{\um}.
In addition to these testbeds, several astronomical nullers have shown on-sky null depths from roughly \num{e-3} to \num{e-1}, from the near- to the mid-infrared.
These include the Multiple-Mirror Telescope (MMT) Nuller\cite{Hinz_1998}, the Keck Interferometer Nuller\cite{Mennesson_2003,Serabyn_2006}, the Palomar Fiber Nuller\cite{Martin_2005,Mennesson_2011} (PFN), the Large Binocular Telescope Interferometer (LBTI) nuller\cite{Defrere_2015,Defrere_2016}, GLINT\cite{Norris_2019,Martinod_2021,Spalding_2024},
and the upcoming Asgard/NOTT\cite{Defrere_2018,Defrere_2022,Laugier_2023}.
A sample of the best nulls achieved on-sky and by testbeds so far is summarized in \cref{fig:nice1_history}.

Despite tremendous progress in the field, no hardware has yet combined stable \num{e-5} nulls with a broadband cryogenic instrument.
The specific high-sensitivity, high-contrast range relevant for LIFE remains experimentally unexplored, and nulling across the full two-octave science bandwidth has never been demonstrated.
Before the space mission can be realized, a ground-based laboratory testbed is thus needed to develop the required technologies and demonstrate the measurement in a realistic environment.

We are building the Nulling Interferometry Cryogenic Experiment\cite{Ranganathan_2024} (NICE) laboratory testbed to fill this technological gap.
NICE will be designed as the first cryogenic nuller focused on high sensitivity and full coverage of the LIFE spectral band, while still achieving the deep and stable nulls that have been demonstrated on previous projects.
NICE will start as a Single-Bracewell nuller\cite{Bracewell_1978}, collecting light from a simulated star with two apertures, to test the capability to reach deep raw nulls and thus suppress photon noise from the star.
Later, it may be extended to a Double-Bracewell design with four beams, representative of the full LIFE beam combiner where a final cross-combiner stage further increases the achievable contrast.

In this paper, we define the laboratory requirements necessary to validate the LIFE beam combiner (\cref{sec:nice1-requirements}), and present the conceptual layout of NICE to meet these requirements (\cref{sec:nice1_methods}).
To prototype the cryogenic instrument, we are building the NICE \enquote{Warm Bench}, an ambient precursor that has already achieved the required null depth with a narrowband \SI{4.7}{\um} laser (\cref{sec:nice1-null_depth_stability}), as well as the required throughput (\cref{sec:nice1-throughput}). 
We discuss our results in \cref{sec:nice1_discussion}, provide an outlook of future work in \cref{sec:nice1_outlook}, and conclude in \cref{sec:nice1_conclusion}.

\section{Requirements}
\label{sec:nice1-requirements}

NICE is a technology demonstrator for the LIFE mission, and the requirements for NICE must be derived such that it could operate as the nulling beam combiner in LIFE.
At the top level, NICE is similar to other high-contrast imaging systems, largely characterized by three key performance metrics: sensitivity, contrast, and spectral bandwidth\cite{Douglas_2018}.
Specifically, NICE should be able to receive two input beams and combine them such that light from a simulated on-axis star is nulled out, while light from a simulated off-axis planet is transmitted to the detector with high efficiency.

Efforts are still ongoing to finalize the LIFE requirements, such that they can systematically be traced back to the LIFE science goals.
A tentative set of requirements on the LIFE measurement performance has already been derived\cite{Dannert_2025_How}, and we choose this as a starting point for deriving the NICE requirements. 
The requirements we set for NICE, summarized in \cref{tab:nice1_reqs}, are intentionally more ambitious than necessary to account for potential upcoming changes in the LIFE requirements.

\begin{table}
    \centering
    \caption[Requirements]{Requirements for NICE to demonstrate the feasibility of the LIFE nulling beam combiner.}
    \vspace{0.2cm}
    \footnotesize
    \begin{tabular}{ll}
         \toprule
         \multicolumn{2}{c}{\emph{During entire observation (with regular re-calibration):}}\\[1.5mm]
         Mean raw null depth &
           \num{2e-5} at \SI{4}{\micro\meter},\\
         & \num{7e-6} at \SI{8}{\micro\meter},\\
         & \num{9e-6} at \SI{10}{\micro\meter}\\
         & \num{1e-4} at \SI{18}{\micro\meter}\\
         Wavelength range & \SIrange{9}{11}{\micro\meter} (requirement)
         \\
         & \SIrange{4}{5}{\micro\meter} (requirement)\\
         & \SIrange{4}{18.5}{\micro\meter} (goal)\\
         Photon conversion efficiency (planet) & \SI{20}{\percent}\\
         \hspace{1em} Throughput & \hspace{1em}\SI{34}{\percent}\\
         \hspace{1em} Detector quantum efficiency & \hspace{1em}\SI{60}{\percent}\\
         Stability without re-calibration & \SI{100}{\s} at least\\
         Calibration time & \SI{10}{\s} (goal)\\
         Cryostat temperature & \SIrange{12}{15}{\kelvin}\\
         \midrule
         \multicolumn{2}{c}{\emph{At planet modulation timescales (band-pass from \SIrange{170}{330}{\micro\hertz}):}}\\[1.5mm]
         Raw null depth RMS & \num{3e-7} at \SI{8}{\um}\\
         \hspace{1em} Optical path-length difference RMS & \hspace{1em}\SI{0.3}{\nano\meter}\\
         \hspace{1em} Intensity RMS (each beam) & \hspace{1em}\SI{0.02}{\percent}\\
         \bottomrule
    \end{tabular}
    \label{tab:nice1_reqs}
\end{table}

\subsection{Raw null depth}
The primary purpose of the nuller is to suppress light from a star through destructive interference, and thus reduce the photon noise it contributes to the measurement.
This capability is qualified by the raw null depth, the intensity ratio of the destructive ($I_-$) to constructive interference ($I_+$), which for a two-beam nuller is defined as
\begin{equation}
\label{eq:nice1_null-depth-definition}
N = \frac{I_-}{I_+} = \frac{I_-}{I_1 + I_2 + 2\sqrt{I_1 I_2}},
\end{equation}
where $I_1$ and $I_2$ are the intensities of the single beams without interference, measured on the science detector at the destructive output.
We put a requirement on the raw null depth because it is easily measurable, while factors that contribute to the null depth, such as differential optical path length and polarization mismatches, are harder to measure.
In practice, there will be two factors that limit the null depth achievable with an on-sky instrument: fundamental astrophysical noise and perturbations by the instrument.

First, fundamental astrophysical noise terms, studied previously in the context of LIFE\cite{Dannert_2022,Hansen_2022}, will leak through even in an ideal nuller.
For example, consider that while an on-axis point source would be perfectly nulled, an extended source, such as a star, will be partially transmitted. This effect is called stellar geometric leakage \cite{Lay_2004} and dominates the noise budget at shorter wavelengths.
At longer wavelengths, the limit is set by the thermal background, such as emission from the local-zodiacal and exozodiacal dust clouds.
Importantly, while some of these background sources can be subtracted in post-processing, even for an ideal nulling interferometer they contribute photon noise, which degrades the planet signal-to-noise ratio (SNR) \cite{Dannert_2022}.

\begin{figure}
    \centering
    \includegraphics[]{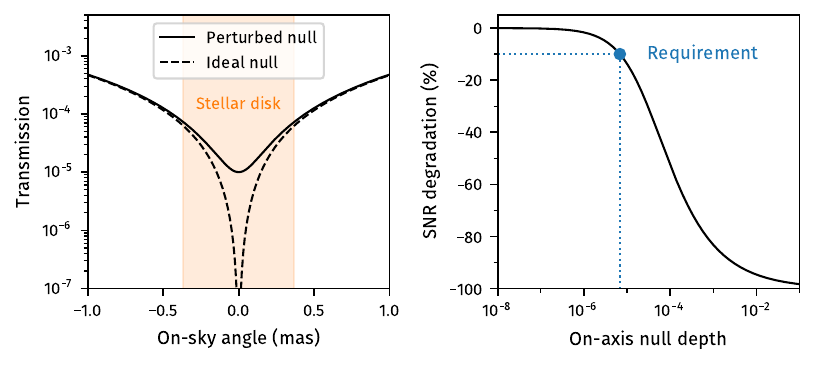}
    \caption[Comparison of an ideal nuller with a perturbed nuller]{
    \textit{Left:} Transmission of a Single-Bracewell nuller at \SI{8}{\um} as a function of on-sky angle for an ideal instrument and a perturbed instrument with degraded visibility. The baseline is optimized for an Earth-twin around a Solar-type star (shaded region) at a distance of \SI{10}{\parsec}.
    The transmission at an on-sky angle of zero is called the on-axis null depth.
    \textit{Right:} The SNR of the instrument degrades as the on-axis null depth worsens because of increased stellar leakage, shown here at \SI{8}{\um} wavelength.
    We set a requirement of at most \SI{10}{\percent} reduction in SNR from instrument errors.
    }
    \label{fig:nice1_derive-on-axis-req}
\end{figure}

Second, a real instrument is additionally subjected to phase and amplitude perturbations, such as residual noise from formation flying or vibrations.
These perturbations, collectively called instrumental errors, deteriorate the quality of the null.
We model this degradation as a decrease of interferometric visibility $V$,
defined as
$V = (I_\text{max}-I_\text{min})/(I_\text{max}+I_\text{min})$,
where $I_\text{max}$ and $I_\text{min}$ are the maximum and minimum intensity that can be achieved on-axis while scanning through fringes.
We schematically show the degradation of the overall null as a function of the on-axis null-depth in \cref{fig:nice1_derive-on-axis-req}.
The on-sky transmission $T$ of the ideal instrument\cite{Dannert_2022} is modeled as
\begin{equation*}
    T(\theta) = \sin^2 \bigg( \frac{\pi L \theta}{\lambda}\bigg),
\end{equation*}
where $\theta$ is the on-sky angle, $L$ is the nulling baseline length, and $\lambda$ is the observing wavelength.
We model the perturbed transmission as a visibility degradation, $T'(\theta) = T(\theta) V + (1-V)/2$.
This allows us to translate between the on-axis null depth as would be measured by NICE, and the null depth over the full field of view as would be measured by LIFE.

To set a requirement on the raw null depth in NICE, we first consider an unperturbed instrument, observing an Earth-twin around a Solar analog, limited only by fundamental astrophysical photon noise sources.
We then find the SNR that this ideal instrument can achieve as a function of wavelength, and demand that perturbations and errors in the beam combiner degrade this ideal SNR by no more than \SI{10}{\percent}.
A similar method was used in Appendix A of Dannert 2022\cite{Dannert_2022}.
This argument is independent of parameters such as integration time and mirror diameter.
The number of \SI{10}{\percent} was chosen because it represents a reasonable trade-off between demands on the instrument and achievable SNR: beyond this number, we enter a regime of strongly diminishing returns, as visible in \cref{fig:nice1_derive-on-axis-req}.
While this is a heuristic requirement, it ensures that fundamental noise sources dominate any contribution from the beam combiner.

The resulting requirement on the null depth in NICE is shown in \cref{tab:nice1_reqs}.
It reaches its deepest value of \num{7e-6} at \SI{8}{\um}, before relaxing by an order of magnitude towards the longer wavelengths.
Improving the null depth beyond this requirement is of limited use, since fundamental astrophysical noise already strongly dominates in this regime.

\subsection{Sensitivity}
\label{ssec:nice1_requirements-sensitivity}

The scientific yield of the LIFE mission depends critically on the capability to collect planet photons with low loss.
We define the photon conversion efficiency (PCE) of LIFE as the number of planet photons detected on the science detector, divided by the number of planet photons entering the primary apertures.
Our reference design\cite{Glauser_2024,Dannert_2025_How} currently assumes four collectors with \SIrange{3}{4}{\m} mirrors, a characterization campaign of five years, a 1:5 ratio between calibration time and science time, and a PCE of \SI{3.5}{\percent}.
The reference PCE will likely increase in the future, but is expected to stay in a range between \SIrange{3.5}{10}{\percent}, depending on the targeted optical design and assumptions on the exoplanet population.

To set a PCE requirement on NICE, we must account for the systems that are represented in NICE, and the systems that are not represented in NICE, such as the primary mirror, the beam transport to the beam combiner, a first stage of correction optics, and the cross-combiner that is used inside a four-beam Double-Bracewell instrument.
Since we can test and characterize NICE in the lab, while the remaining system has not been experimentally verified, we chose to put a conservative \SI{20}{\percent} PCE requirement on NICE.
Optimizing sensitivity even beyond this requirement is an ambition strongly worth pursuing, as
gains in throughput can relax other parameters and reduce mission cost.\cite{Ireland_2024}.
We further break the PCE requirement down into a requirement on the throughput of the optics (\SI{34}{\percent}) and a requirement on the quantum efficiency (QE) of the detector (\SI{60}{\percent}).
A preliminary throughput budget is shown in \cref{sec::nice1_throughput-budget}.

\subsection{Null stability}
\label{ssec:nice1_stability}

\begin{figure}
    \centering
    \includegraphics[]{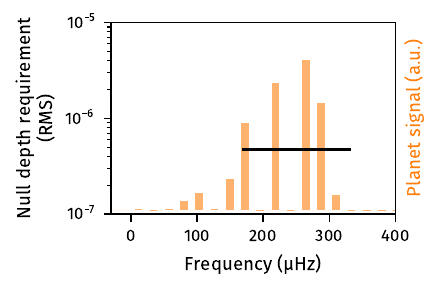}
    \caption[Requirement on the null depth power spectrum.]{
    Requirement on the null depth power spectrum at \SI{8}{\um}.
    Assuming one rotation of the array per day, NICE requires a stability of the null depth of \num{6e-7} RMS from \SIrange{170}{330}{\micro\Hz} (black line) for a Single-Bracewell interferometer to act as the beam combiner in LIFE. The expected spectral signature of a planet is also shown.}
    \label{fig:nice1_stability-req}
\end{figure}

Slow drifts of the instrument lead to a measurement that could be mistaken for a planet signal, similar to how speckles can be mistaken for companions in single-aperture high-contrast imaging, and thus a requirement on the stability of the null depth is critical to achieve a confident measurement.\cite{Dannert_2025_Consequences}
More specifically, LIFE records time-series of spectra, and the spatial resolution of LIFE --- i.e. the capability to locate a planet and to distinguish two planets in the same system --- is achieved through a rotation of the interferometric baselines over time.
In the Fourier transform of this time series, we know that planet signals can only occur at specific harmonics of the rotation period of the array, as shown in \cref{fig:nice1_stability-req}.
As a consequence, any noise in the null depth within this frequency range could be mistaken for a planet signal, and result in a false positive detection.
It has been derived\cite{Dannert_2025_How} that a noise level of \SI{0.6}{\nm} RMS in OPD and \SI{0.04}{\percent} RMS in beam intensity within \SIrange{170}{330}{\micro\hertz} (assuming one baseline rotation per day) can be allowed for LIFE to achieve its science goals.
This corresponds to a null depth RMS of \num{6e-7} RMS at \SI{8}{\um} at these frequencies.
To account for noise sources not modeled in NICE, such as formation flight and beam transfer to the combiner, we impose the stricter requirements in \cref{tab:nice1_reqs} on the testbed, corresponding to a \num{3e-7} null depth RMS.

\subsection{Duty cycle and automatic recalibration}
\label{ssec:nice1_duty-cycle-req}

To meet the constraints on mean null depth and null depth stability, we expect regular calibrations during science observations.
With the assumed 5:1 ratio between science observations and calibration for LIFE, we set the requirement that at most \SI{10}{\percent} of operation time may be used for calibration procedures in NICE, accounting for NICE calibrating approximately half the number of beams as LIFE.

Previous benches, such as PERSEE \cite{Lozi_2010} and the Planet Detection Testbed \cite{Martin_2012}, used a dithering technique to find the reference point of the OPD and tip/tilt of the fiber injection: measuring either side of the current position and using prior information on the null to find the new reference point \cite{Gabor_2008_Stabilising}.
To properly constrain the phase and amplitude of both beams, ten measurements must be made: eight for the pointing and shear (or lateral position) of each beam, and two for the OPD. 
From Appendix A of Hansen et al.\ (2023) \cite{Hansen_2023}, in order to be dominated by emission from zodiacal dust across the LIFE bandbass, one needs to integrate the dither on the order of \SIrange{0.5}{1}{s}.
Including settling time of the various piezo stages, this results in a calibration on the order of \SI{10}{\s}.
Thus, the NICE measurement must be stable for at least \SI{100}{\second} to meet the requirement on duty cycle, and we target a calibration time of \SI{10}{\s} with realistic stellar flux levels.

\section{Methods}
\label{sec:nice1_methods}

The primary goal of NICE is to demonstrate the nulling beam combiner which lies at the heart of the LIFE beam combiner spacecraft.
The optical system of NICE thus has to be representative of the space hardware, which demands a broadband, cryogenic, high-throughput instrument.

\subsection{Optical setup}

\begin{figure}
    \centering
    \includegraphics[width=\textwidth]{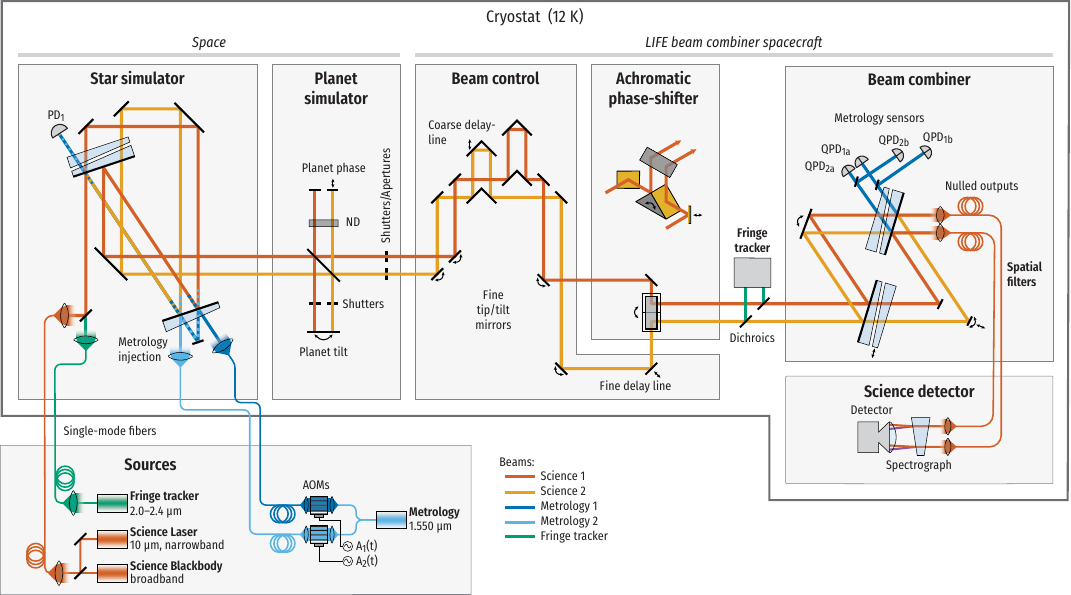}
    \vspace{1em}
    \caption[Preliminary optical diagram of NICE]{Preliminary optical diagram of NICE. QPDs are quadrant photodiodes for the metrology, and PDs are regular photodiodes. The electrical signals $A_1$ and $A_2$ are amplitude-modulated drive signals for the acousto-optic modulators (AOMs)}
    \label{fig:nice1_cryo-diagram}
\end{figure}

\begin{figure}
    \centering
    \includegraphics[width=0.8\textwidth]{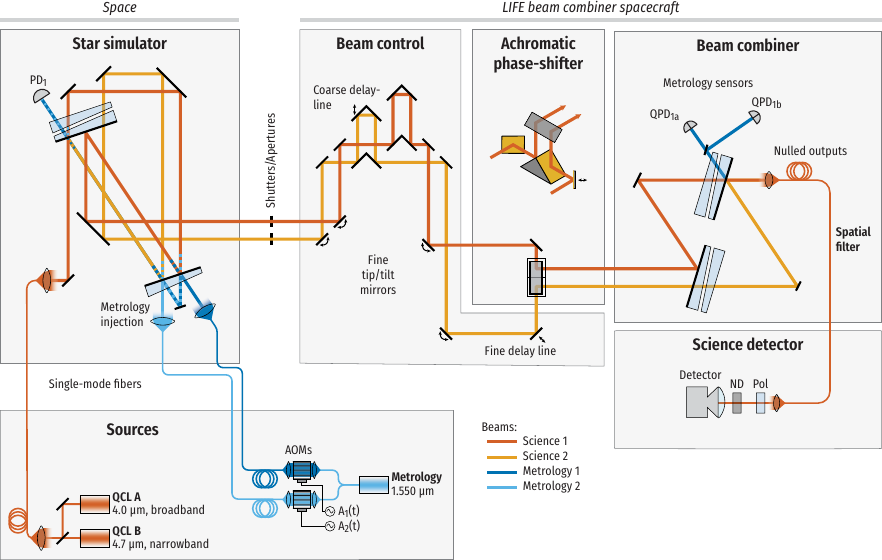}
    \vspace{1em}
    \caption[Optical diagram of the Warm Bench]{Optical diagram of the Warm Bench, a precursor for NICE at ambient conditions, as it was used for the measurements.}
    \label{fig:nice1_warm-diagram}
\end{figure}

\begin{figure}
    \centering
    \includegraphics[width=0.9\textwidth]{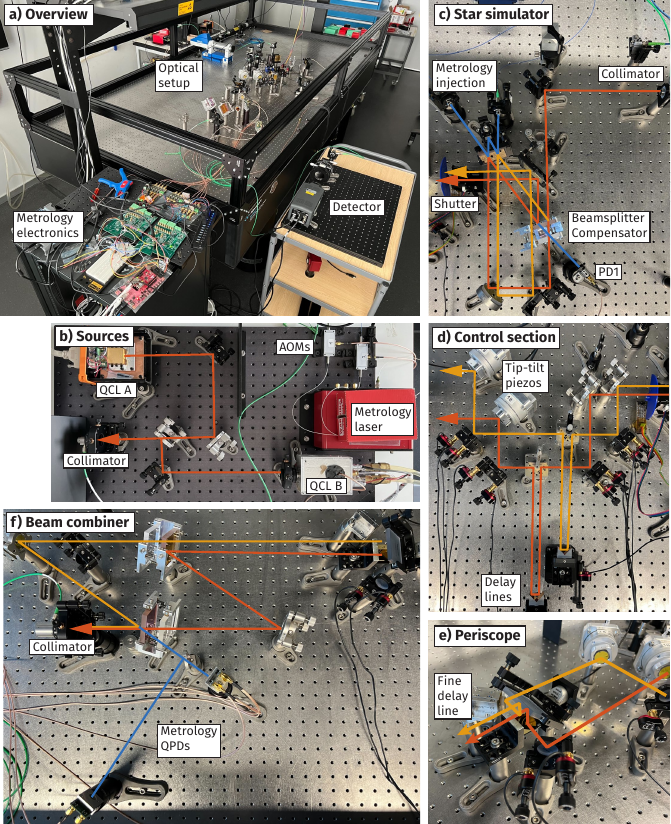}
    \vspace{0.2cm}
    \caption[Pictures of the Warm Bench]{Subsystems of the Warm Bench. a) Overview b) Sources c) Star simulator d) Control section e) Periscope f) Beam combiner.}
    \label{fig:nice1_warm-pictures}
\end{figure}

We show two optical diagrams in this section:
the preliminary optical diagram of the cryogenic NICE bench (\cref{fig:nice1_cryo-diagram}), and the optical diagram of the NICE Warm Bench (\cref{fig:nice1_warm-diagram}, \cref{fig:nice1_warm-pictures}) in its current state, which is the precursor experiment at ambient conditions, used for all measurements that follow.
We focus our description on the cryogenic setup, and note the differences with the warm setup wherever they occur.
To reduce differential chromatic and polarization errors in the science beams, we designed the setup such that it is symmetric, i.e. the two science beams undergo the same number of transmissions and reflections.
All mirrors will be coated with unprotected gold for the cryogenic setup, and protected gold is used on the Warm Bench.
Beam splitters use a \ce{CaF2} substrate on the Warm Bench, and \ce{ZnSe}, \ce{KBr}, or other similarly mid-IR transmissive substrates will be used for longer wavelengths in the cryogenic setup.

\subsubsection{Sources}
Since this is a laboratory testbed, we have to simulate the star.
We aim to eventually cover the full LIFE science band of \SIrange{4}{18.5}{\um}, which will likely be implemented as a broadband thermal blackbody.
Multiple narrowband sources will be used for prototyping and characterization, as a blackbody source will not produce enough power for easy measurements.
In addition to the science sources, we foresee separate sources for a fringe tracker and a laser metrology system, explained in more detail in \cref{ssec:nice1_metrology-control}.

On the Warm Bench, we chose to start with the short end of the science band, where requirements on geometric stability are most stringent and the thermal background of the lab is lower.
We currently use two quantum-cascade lasers (QCLs) to simulate a star for the setup:
One is a \SI{4.0}{\um}, \SI{100}{\milli\watt} off-the-shelf laser with coherence length on the order of $\approx \SI{100}{\um}$, and is used to find the central dark fringe, but not for nulling measurements, as it is a pulsed source.
(This source has been recently replaced with a continuous-wave source for future broadband nulling experiments.)
The other source is a \SI{4.7}{\um}, \SI{100}{\milli\watt} continuous-wave laser with coherence length $> \SI{10}{\centi\meter}$, developed in-house, and used as a narrowband source for nulling.
The sources are located on a separate table, and are fed into the star simulator with single-mode fibers.
Both lasers can be injected into the setup simultaneously, using a beam-splitter to couple into the same fiber.
While we expect top-hat shaped beams from the LIFE collectors, all sources in the current setup have a Gaussian profile.

\subsubsection{Star simulator}
The single beam from a laser is split into two, representing beams received from two of the collector spacecraft.
Because of its inherent symmetry, we use a modified Sagnac interferometer to simulate the star, where each beam undergoes an equal number of transmissions and reflections, and therefore accumulates almost identical chromatic and polarization errors.
To avoid ghosts and chromatic errors, all beam-splitters that the science beams traverse are wedged and equipped with compensator plates, such that the length of glass seen by each beam is identical.
In a future extension, we may require four identical output beams for a Double-Bracewell interferometer, hence requiring a multi-stage system.

\subsubsection{Planet simulator}
To demonstrate that a star is nulled while a planet is still transmitted, we need to simulate a second source.
A real planet will have both an off-axis angle, as well as an additional phase-shift compared to the star.
As a simpler starting point, part of the stellar light can be split off after the star simulator, tilted slightly, and phase shifted, such that the desired planet signal can be simulated.
Because of the large optical path length in the planet simulator, the planet light would be incoherent with the stellar light, as required.
However, this does not simulate a realistic planet spectrum, which would be much colder than that of the star.
A possible extension is to use a second star simulator with a different blackbody source to produce a more realistic planet.
At the output of the planet simulator, we implement a set of beam-shutters that can block either or both beams from entering the setup, which we use for intensity calibration and for background measurements.
The planet simulator is currently not implemented on the Warm Bench.

\subsubsection{Correction stage}

The correction stage is the first subsystem in the optical train of NICE that will also be part of the LIFE beam combiner spacecraft, and as such is also the starting point for throughput measurements.
We do not model the collector mirrors or the beam transport between collectors and combiner in NICE.
The purpose of this section is to control the OPD between the beams and their alignment with respect to the spatial filter at the end of the setup.
Two piezo tip/tilt stages per beam allow control of shear and pointing of each beam.
A slow delay line with long stroke, based on a slip-stick piezo actuator, is used for the initial OPD scan to find the zero-OPD fringe.
A fast piezo delay line stabilizes the OPD during nulling.
The fast delay line and the tip/tilt stages are in closed-loop with the metrology system and the fringe tracker.

\subsubsection{Achromatic phase-shifter}
The achromatic phase shifter introduces a relative phase-shift of $\pi$ between the beams, such that destructive interference is achieved in the beam combiner.
It is implemented as a periscope \cite{Rabbia_2003}, which we chose because it is inherently achromatic if well-aligned, as it only uses reflections to achieve the phase shift.
Alternative methods, based on transmissive optics such as phase plates, are not feasible for our broad wavelength-range requirement without also requiring further correcting subsystems, such as an adaptive nuller\cite{Lay_2003,Peters_2008}, or splitting the wavelength range into smaller channels.

\subsubsection{Beam combiner and spatial filter}
A modified Mach-Zehnder interferometer\cite{Serabyn_2001} combines the two beams.
It offers the advantage of symmetry and thus achromaticity, but it does produce duplicate outputs: two outputs with constructive interference, and two outputs with destructive interference.
Only one destructive output is currently available on the Warm Bench, since a second fine delay-line would be required to achieve nulls at both outputs simultaneously.
Both outputs will be obtained in the future to demonstrate the required system sensitivity.

The beams from the nulled output of the beam combiner, now overlapping and with a $\pi$ phase difference, are coupled into a single-mode spatial filter.
This relaxes the requirements on wavefront-quality, such that the operation of nulling is essentially reduced to matching intensity and phase for each wavelength and polarization mode of the filtered beams.
However, since coupling into the fiber depends on the position and angle of the incoming beams, the throughput of the system now strongly depends on alignment, and leakage of higher-order modes will still be present, although highly attenuated.
On the Warm Bench, we currently use an off-the-shelf \SI{5}{\meter} single-mode \ce{InF3} fiber, with a single-mode cutoff wavelength of \SI{3.2}{\micro\meter}, and off-axis parabolas as collimators.

\subsubsection{Detector}
The two outputs from the spatial filter are collimated and spectrally dispersed with a prism.
On the Warm Bench, we use an off-the-shelf InSb detector-array at \SI{77}{\kelvin}, located outside of the enclosure to reduce the coupling of vibrations from its cryocooler to the interferometer.
In the cryogenic setup, detector options include a \ce{HgCdTe} array (although such an option may have limited use at the longest wavelengths), or a superconducting detector such as a kinetic inductance detector (KID), which is currently under development for efficient mid-infrared operation. \cite{Ras_2024}

\subsection{Throughput budget}
\label{sec::nice1_throughput-budget}

\begin{table}
    \centering
    \scriptsize
    \caption[Throughput budget]{Preliminary throughput budget for LIFE and NICE.
    Data from off-the-shelf components were used where available, and estimates given in other cases.
    Efficiency numbers are for a single component in the detailed breakdown, and total efficiency for bold numbers.
    All mirrors are assumed coated with raw gold ($R=\SI{98.8}{\percent}$) unless otherwise noted.
    }
    \label{tab:throuhput-budget}
    \vspace{0.2 cm}
    \begin{tabular}{lS[table-format=2.1]S[table-format=2.1]ll}
        \toprule
        \textbf{Component} & \textbf{Qty.} & \multicolumn{2}{c}{{\textbf{Efficiency (\%)}}} & \textbf{Notes} \\
        \cmidrule(lr){3-4} & & {\SI{4}{\um}} & {\SI{10}{\um}} & \\
        \midrule
        \textbf{Collectors} & & \bfseries 94.1 & \bfseries 94.1 & \\
        Off-axis Cassegrain & 3 & 98.0 & 98.0 & three reflections, protected gold\\[0.2 cm]
        \textbf{Receiving Stage} & & \bfseries 64.3 & \bfseries 64.3 & \\
        Steering mirrors & 2 & 98.8 & 98.8 & two tip/tilt mirrors\\
        Polarization compensation & 2 & 98.8 & 98.8 & two reflections\\
        Beam compression & 2 & 98.8 & 98.8 & reflective Keplerian\\
        DM & 1 & 98.0 & 98.0 & protected gold, no window\\
        PIAA & 2 & 98.8 & 98.8 & two aspherical mirrors\\
        Dichroics & 2 & 85 & 85 & JWST Mid-IR dichroics at \SI{7}{\kelvin}, typical\cite{Hawkins_2007}\\[0.2 cm]
        \textbf{Control Section} & & \bfseries 73.5 & \bfseries 73.5 & \\
        Long delay line & 4 & 98.8 & 98.8 & trombone, four reflections\\
        Fine delay line & 1 & 98.8 & 98.8 & fold mirror, one reflection\\
        Fine steering mirrors & 2 & 98.8 & 98.8 & two tip/tilt mirrors\\
        Intensity control & & 80 & 80 & via shear/pointing into spatial filter \\[0.2 cm]
        \textbf{Adaptive Nuller (optional)} & &\bfseries 84.7 &\bfseries 88.2 & \\
        Dispersing prism & 2 & 97 & 98 & ZnSe with AR-coating (Thorlabs \enquote{E4})\\
        Wollaston prism & 2 & 97 & 98 & estimate for CdSe with AR coating ($n \approx 2.4$, similar to ZnSe)\\
        Parabola & 2 & 98.8 & 98.8 & \\
        DM & 1 & 98.0 & 98.0 & Protected gold\\[0.2 cm]
        \textbf{Achromatic Phase Shifter} & &\bfseries 97.6 &\bfseries 97.6 & \\
        Periscope & 2 & 98.8 & 98.8 & two reflections, can use as fold mirror\\[0.2 cm]
        \textbf{Beam Combiner (MMZ)} & &\bfseries 89.6 &\bfseries 84.7 & \\
        Mirror & 1 & 98.8 & 98.8 & \\
        Beamsplitter reflection & 1 & 42.0 & 42.0 & \ce{CaF2} at \SI{4}{\um}, \ce{ZnSe} at \SI{10}{\um} \\
        BS transmission + compensator & 1 & 54.0 & 51.0 & same as above \\
        Number of total beams & & {$\times 4$} & {$\times 4$} & Two destructive outputs, two beams each\\[0.2 cm]
        \textbf{Spatial Filter} & &\bfseries 70.3 &\bfseries 54.5 & \\
        Coupling optics & 1 & 98.8 & 98.8 & Off-axis parabola\\
        Fresnel loss at facet & 1 & 95 & 95 & estimate\\
        Strehl & & 85 & 85 & $\approx \SI{650}{\nm}$ wavefront error at \SI{10}{\um}\\
        Propagation loss & 1 & 89.1 & 69.2 & \SI{5}{\m} \ce{InF3} at \SI{4}{\um}, \SI{20}{\cm} chalcogenide single-mode\cite{Ksendzov2007} at \SI{10}{\um}\\
        Collimation optics & 1 & 98.8 & 98.8 & Off-axis parabola \\[0.2 cm]
        \textbf{Cross-combiner} & &\bfseries 89.0 &\bfseries 86.2 & \\
        Fold mirrors & 2 & 98.8 & 98.8 & \\
        Phase plate & 1 & 95 & 95 & \ce{MgF2} at \SI{4}{\um} \\
        Beam splitter ($R+T$) & 1 & 96 & 93 & same as in beam combiner, both outputs used \\[0.2 cm]
        \textbf{Spectrograph} & & \bfseries 95.8 &\bfseries 96.8 & \\
        Dispersing prism & 1 & 97 & 98 & ZnSe with AR-coating, same as adaptive nuller\\
        Imaging optics & 1 & 98.8 & 98.8 & Off-axis parabola\\[0.2 cm]
        \textbf{Detector} & &\bfseries 70 &\bfseries 60 & \\
        Quantum efficiency &  & 70 & 60 & Hawaii 2RG at \SI{4}{\um}, JWST MIRI at \SI{10}{\um}\\
        \midrule
        \textbf{Total throughput} & & \bfseries 19.8 &\bfseries 14.8 & \\
        \textbf{Total PCE} & & \bfseries 13.8 & \bfseries 8.9 & \\
        \bottomrule
    \end{tabular}
\end{table}

To motivate how much of the \SIrange{3.5}{10}{\percent} range for the PCE requirement of LIFE to allocate to subsystems in NICE and the Warm Bench, and to motivate that this is feasible, an approximate preliminary throughput budget is shown in \cref{tab:throuhput-budget}.

For more information on the optical design of the LIFE mission and its optical diagram, we refer to the reference design\cite{Glauser_2024}.
Main uncertainties in this budget are losses from spatial filtering, detector efficiency, and efficiencies of some of the optical substrates and coatings.
The Warm Bench includes the control section, the achromatic phase shifter, the beam combiner, and the spatial filter, which together have a budgeted throughput of \SI{45.2}{\percent}, or \SI{22.6}{\percent} if only one of the two outputs of the beam combiner is used.
The remaining systems, which are the collector, the receiving stage, an optional adaptive nuller, the cross-combining stage, and a spectrograph, have a budgeted throughput of \SI{43.7}{\percent}.

Note that some of the estimates in the budget are conservative, since for example the adaptive nuller is an optional compensating subsystem, and if it were to be included, other subsystems would likely be simplified or removed, such as some dichroics and the MMZ.
This is discussed in more detail in \cref{ssec:compensation}.
Also, a \SI{20}{\percent} loss is allocated to intensity control, which provides margin for compensating intensity mismatches in the optical chain.
Most of the chosen coatings are off-the-shelf or legacy options, and not optimized for LIFE's science bandpass.

In comparison with the requirement range of \SIrange{3.5}{10}{\percent} PCE from \cref{ssec:nice1_requirements-sensitivity}, the budgeted \SI{8.9}{\percent} PCE at \SI{10}{\um} has approximately \SI{150}{\percent} margin to the requirement of \SI{3.5}{\percent}, while not quite achieving the more ambitious end of \SI{10}{\percent} PCE.

\subsection{Error budget}
\label{ssec:nice1_error-budget}

\begin{figure}
    \centering
    \includegraphics[width=0.7\textwidth]{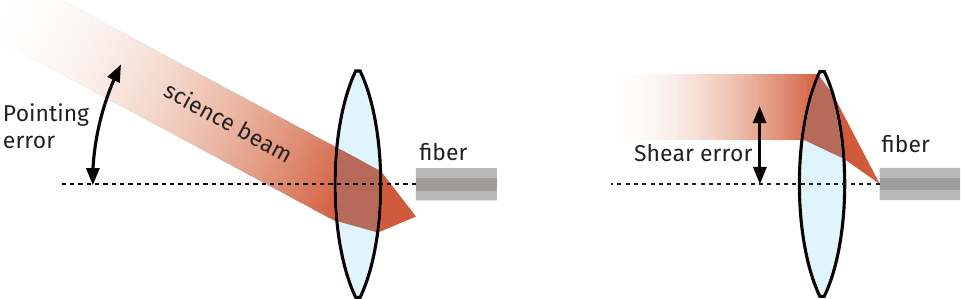}
    \vspace{1em}
    \caption[Definition of pointing and shear]{Definition of pointing and shear. Pointing refers to the angular error of the science beam before the collimator, and shear to the lateral displacement of the science beam before the collimator.}
    \label{fig:nice1_pointing-shear}
\end{figure}

The requirement on null depth in NICE from \cref{tab:nice1_reqs} can be further broken down into the equal-weight error budget in \cref{tab:nice1_error-budget}.
The contribution of individual errors in the system to the mean null depth, derived in \cref{sec:nice1_derive-error-budget}, is
\begin{equation}
    \label{eq:nice1_null-dynamic}
    \mean{N}
    = \frac{1}{4} \Big( \mean{\delta\phi}^2 + \sigma_{\delta\phi}^2 + \frac{1}{4}\mean{\delta\phi_\text{sp}}^2 + \mean{\delta I}^2 + \sigma_{\delta I}^2 + \frac{1}{4} \mean{\delta I_\text{sp}}^2 \Big),
\end{equation}
where $\mean{N}$ is the mean null depth at a single wavelength and spatial mode, $\delta\phi$ is the phase error between the beams, $\delta I = \abs{I_1 - I_2}/(I_1 + I_2)$ is the intensity mismatch between beams, and terms with an sp-subscript indicate differential errors between the s and p polarization modes.
Angled brackets $\langle \dots \rangle$ refer to the mean of a quantity over time, and $\sigma$ refers to the sample standard deviation over time.
More detailed definitions of the terms as well as the derivation of the equation are in \cref{sec:nice1_derive-error-budget}.
This model assumes an ideal single-mode spatial filter, coherent beams, and that all polarization errors are static in time.
The intensity mismatch depends on the coupling efficiency of the beams into the single-mode spatial filter, and can be further broken into contributions from pointing error, shear error (see \cref{fig:nice1_pointing-shear}), and higher-order terms such as lab seeing and surface errors.
The higher-order contributions are reported as fluctuations in intensity, and can be related to the wavefront error or Strehl ratio.

\begin{table}
    \centering
    \caption[Error budget]{Equal-weight error budget for NICE to achieve the required mean null depth $\mean{N}$.
    Assumptions: Gaussian beams with \SI{6}{\mm} diameter, static polarization errors, and a single-mode spatial filter.
    For static errors, we put requirements on the mean value of phase $\mean{\delta\phi}$ and intensity errors $\mean{\delta I}$, while for dynamic errors, we put requirements on their sample standard deviations $\sigma_{\delta\phi}$ and $\sigma_{\delta I}$.
    The most stringent demands on phase errors are at \SI{4}{\um}, and the most stringent demands on intensity errors are at \SI{8}{\um}.
    }
    \vspace{0.2cm}
    \footnotesize
    \begin{tabular}{lllll}
         \toprule
         Parameter & \multicolumn{4}{c}{Wavelength} \\
         \cmidrule(lr){2-5} & \SI{4}{\um} & \SI{8}{\um} & \SI{12}{\um} & \SI{18}{\um} \\
         \midrule
         Required $\mean{N}$ & \num{2.0e-5} & \num{6.8e-6} & \num{1.8e-5} & \num{1.0e-4}\\[0.5em]
         \emph{Static errors:}\\
         \hspace{1em} $\mean{\delta\phi} $ & \SI{2.3}{\nano\meter} & \SI{2.7}{\nano\meter} & \SI{6.7}{\nano\meter} & \SI{24}{\nano\meter}\\
         \hspace{1em} $\mean{\delta I}$ & \SI{0.37}{\percent} & \SI{0.21}{\percent} & \SI{0.35}{\percent} & \SI{0.83}{\percent}\\
         \hspace{1em} $\mean{\delta I_\text{sp}}$ & \SI{0.74}{\percent} & \SI{0.43}{\percent} & \SI{0.70}{\percent} & \SI{1.7}{\percent}\\
         \hspace{1em} $\mean{\delta\phi_\text{sp}}$ & \SI{4.7}{\nm} & \SI{5.4}{\nm} & \SI{13}{\nm} & \SI{48}{\nm}\\[0.5em]
         \emph{Dynamic errors:}\\
         \hspace{1em} $\sigma_{\delta\phi}$ & \SI{2.3}{\nano\meter} & \SI{2.7}{\nano\meter} & \SI{6.7}{\nano\meter} & \SI{24}{\nano\meter}\\
         \hspace{1em} $\sigma_{\delta I}$ & \SI{0.37}{\percent} & \SI{0.21}{\percent} & \SI{0.35}{\percent} & \SI{0.83}{\percent}\\
         \hspace{2em} $\sigma_\text{pointing}$ & \SI{16}{\micro\radian} & \SI{25}{\micro\radian} & \SI{47}{\micro\radian} & \SI{110}{\micro\radian}\\
         \hspace{2em} $\sigma_\text{shear}$ & \SI{204}{\micro\meter} & \SI{155}{\micro\meter} & \SI{199}{\micro\meter} & \SI{307}{\micro\meter}\\
         \hspace{2em} $\sigma_\text{higher order}$ & \SI{0.18}{\percent} & \SI{0.11}{\percent} & \SI{0.17}{\percent} & \SI{0.42}{\percent}\\
         \bottomrule
    \end{tabular}
    \label{tab:nice1_error-budget}
\end{table}

\subsection{Metrology and control system}
\label{ssec:nice1_metrology-control}

For NICE to achieve the stability budgeted in \cref{ssec:nice1_error-budget}, we developed a metrology and control system\cite{Birbacher_2024} to monitor changes in OPD, pointing, and shear, and apply corrections to the fine delay line and the tip/tilt-stages.
The metrology system currently implemented on the Warm Bench is not representative of the flight configuration, since we also have to monitor the star simulator and the planet simulator in the testbed, though a similar concept may be used for the LIFE beam combiner internally.
The metrology laser is a \SI{100}{\milli\watt}, \SI{1.55}{\um} solid-state laser, used for both OPD and beam position measurements.
The OPD metrology is based on a heterodyne interferometer with a phase-meter, while beam positions are measured using the same beams with an intensity-modulation scheme and quadrant photodiodes (QPDs).
The metrology beams are overlaid with the science beams in the star simulator, and any error measured in the metrology beams is used as a proxy for errors in the science beams, though drifts and non-common-path errors may occur.
One pair of metrology beams propagates backwards and is measured at $\text{PD}_1$, while the other pair of metrology beams propagates forwards, and is measured at the QPDs at the constructive output of the beam combiner.
The system achieves best-case OPD control within \SI{0.7}{\nm} RMS residual metrology error in closed-loop operation with a $-\SI{3}{\dB}$ control bandwidth of $\approx \SI{20}{\Hz}$, and a measurement bandwidth of \SI{1}{\kHz}.
Shear is controlled in closed loop, but at a much slower bandwidth ($\approx \SI{2}{\Hz}$), with $\approx \SI{0.2}{\um}$ RMS residual metrology error with a \SI{100}{\Hz} measurement bandwidth.
Pointing is also measured at \SI{100}{\Hz} bandwidth, but pointing control is not yet implemented.
Pointing drifts are slow enough not to contribute significantly to current null measurements.

While the laser metrology can see the full beam path relevant for NICE, LIFE will need a different system to monitor pathlengths between the spacecraft and the star.
We foresee a fringe-tracker in the cryogenic NICE setup to demonstrate measuring drifts in OPD at timescales of \SI{0.1}{\s} or longer with realistic flux levels.
The OPD control loop will then use fringe tracker measurements at slower frequencies, and metrology measurements for high-frequency control.
The fringe tracker will operate in the astronomical K-band ($\approx \SIrange{2.0}{2.4}{\um}$) from a broadband source, representing the wavelengths of the star below the science band.
An integrated photonic device similar to the GRAVITY fringe tracker\cite{Perraut_2018,Lacour_2019} could be used, and is under development at ETH Zürich.

\subsection{Alignment and measurement procedure}

The setup is initially aligned by placing the components on the table in approximately the nominal position, and then fine-tuned with an autocollimator and an auxiliary Michelson interferometer to ensure good angular precision.
A visible alignment laser that is co-aligned with the science lasers is used for this process.
Once fringes are visible at the destructive output of the beam combiner, the spatial filter is placed at their location, and alignment is tuned to maximize coupling into the spatial filter.
To find zero OPD, we use the pulsed broadband laser with a coherence length on the order of \SI{100}{\um}, and scan the rough delay line until the global minimum of the intensity pattern is reached.
At that point, we switch to the narrowband laser, as it is a continuous rather than a pulsed source.

With the beam shutters, we chop between the beams and adjust the tip/tilt mirrors slightly to balance their intensities.
We open the beam shutters such that both beams interfere, remove the \SI{0.1}{\%} ND filter in front of the detector, and start the closed-loop control system to stabilize the beams.
A short iterative process of optimizing the null depth begins, which consists of manually scanning the OPD setpoint and the shear setpoints until the spot is as dark as possible, and the measurement of the null depth is started.
After the null depth measurement, the beam control system is switched off, the ND filter is re-introduced, and a time-series of the intensity of each beam is measured to derive the maximum constructive intensity.
For all recordings, an area $\approx$ 64 x 64 pixels on the detector is recorded at $\approx \SI{1000}{\kHz}$ sampling rate, with an exposure time of \SI{0.15}{\ms}.

To analyze the data, all recordings are background-subtracted and corrected for the ND filter.
The flux in a circular aperture is measured as a function of time.
For the intensities $I_1$ and $I_2$, the median of this time series is used as a proxy for the maximum constructive intensity.
Note that the beams from the constructive output cannot be used for this, since they do not couple into the same spatial filter, and thus do not experience the time-varying coupling-efficiency into the single-mode fiber as the destructive output.
The null depth is finally calculated according to \cref{eq:nice1_null-depth-definition}.

\section{Results}

\subsection{Null depth and stability}
\label{sec:nice1-null_depth_stability}
\begin{figure}
    \centering
    \includegraphics{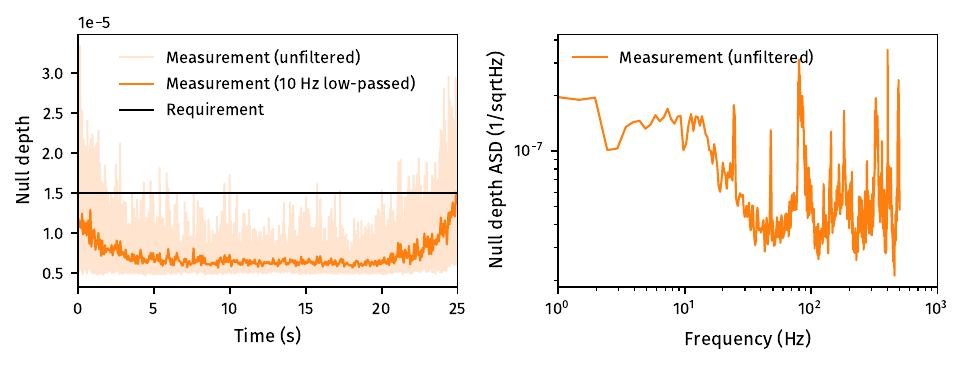}
    \caption[Measured null depth]{\emph{Left:} The measured null depth meets the requirement for \SI{25}{\s}, before a drift in OPD occurs that was not visible to the metrology system.
    \emph{Right:} Amplitude spectral density of the same measurement.}
    \label{fig:nice1_null-time-series}
\end{figure}

\Cref{fig:nice1_null-time-series} shows a measurement of a \num{7.2e-6} mean null depth over \SI{25}{\second}, using the narrowband \SI{4.7}{\um} laser source, and with a linear polarizer at the output.
This meets our requirement of \num{1.5e-5} at \SI{4.7}{\um}.
There is a drift towards the end of the measurement, and the stability of the null is limited to $\approx \SI{25}{\s}$.
The result was repeatable, with a mean null depth of $< \num{e-5}$ achieved in multiple sessions.

\subsection{Throughput}
\label{sec:nice1-throughput}

The starting point for the throughput measurement in NICE is at the beginning of the beam control section, since the star simulator and the planet simulator are not part of LIFE.
We used the \SI{4.0}{\um} QCL as a source, and we set the shutter to block one of the science beams at a time, to measure the throughput of each beam independently without interference effects.
We used a sensitive mid-IR wavefront sensor to measure the relative power loss of each section, and we used a less sensitive thermal power meter to verify our measurements up to the MMZ, where the power meter was no longer sufficiently sensitive.

\begin{table}
\centering
    \caption[Throughput measurement]{Throughput measurement of NICE at \SI{4.0}{\um} with best-effort fiber coupling. Some additional losses occur when balancing intensities between the two science beams. Only one of the two MMZ outputs is currently implemented, and the requirement is thus only half that of the two-output system.}
    \vspace{0.2cm}
    \footnotesize
    \begin{tabular}{
        l
        S[table-format=2.1+-1]
        S[table-format=2.1]
        S[table-format=2]
        c
        }
         \toprule
         \multirow{2}{*}[-1mm]{{Location}} &
         \multicolumn{4}{S}{{Throughput (\%)}}\\
         \cmidrule(lr){2-5}
         & {Measured} & {Budget} & {Required}  & Status\\
         \midrule
        Control section (no $\delta I$-control) & 90 \pm 2 & 91.9 & {---} & {---} \\
        Achromatic phase-shifter & 95 \pm 2 & 97.6 & {---} & {---} \\
        Beam combiner (single output) & 40 \pm 2 & 44.8 & {---} & {---} \\
        Spatial filter & 63 \pm 2 & 70.3 & {---} & {---} \\
        \midrule
        Total & 22 \pm 1 & 28.2 & 17 & $\checkmark$\\
        \bottomrule
    \end{tabular}
    \label{tab:nice1_throughput}
\end{table}

\Cref{tab:nice1_throughput} shows the measured throughput of each section of the setup, as well as the throughput of the full setup.
In the numbers presented here, we assume constructive interference for planet light, which is also how the requirement is derived.
We measure a throughput of \SI{22\pm 1}{\percent}, which meets the requirement of \SI{17}{\percent} for a single output of the MMZ.
With both outputs of the MMZ active, we would require a \SI{34}{\percent} throughput.

The measured value is lower than the budgeted value of \SI{28.2}{\percent}, mostly due to the use of less efficient protected gold mirror coatings on the Warm Bench, and likely sub-optimal coupling into the spatial filter.
If protected gold coatings ($R=\SI{98.0}{\percent}$) are used in the budget, the resulting expected throughput of \SI{23.6}{\percent} is consistent with the measurement.

\section{Discussion}
\label{sec:nice1_discussion}

\subsection{Mean null depth}

\begin{figure}
    \centering
    \includegraphics{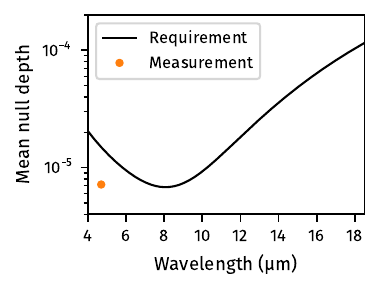}
    \caption[Null depth over wavelength]{Comparison of measured null depth at \SI{4.7}{\um} with the required null depth over the mission wavelength range. The measured null depth meets the requirement with a 2x margin, and would almost meet the most stringent requirement at \SI{8}{\um}.}
    \label{fig:nice1_null-meas-vs-req}
\end{figure}

The measured mean null of \SI{7e-6} is compliant with the requirement at \SI{4.7}{\um}, and is close to even the deepest requirement of \SI{6.8e-6} at \SI{8.0}{\um}, as shown in \cref{fig:nice1_null-meas-vs-req}.

\begin{table}
\centering
    \caption[Null depth residual error analysis]{The measured residual errors during the null depth measurement meet the error budget for the required null depth. Note that all measurements except for null depth were obtained with the metrology system, so potential offsets or drifts between metrology and science are possible. The pointing measurement was not yet implemented at the time when the nulls were recorded, the values reported here were measured later and are typical for an open enclosure.}
    \vspace{0.2cm}
    \footnotesize
    \begin{tabular}{llllc}
         \toprule
         Parameter & Measured & Error budget & Status \\
         \midrule
         Mean null depth & \num{7.17e-6} & \num{1.5e-5} & $\checkmark$\\
         $\sigma_{\delta\phi}$ & \SI{1.2}{\nm} & \SI{2.4}{\nm} & $\checkmark$ \\
         $\sigma_\text{shear}$ & \SI{0.17}{\um} & \SI{203}{\um} & $\checkmark$ \\
         $\sigma_\text{pointing}$ & $\approx \SI{10}{\micro\radian}$ & \SI{19}{\micro\radian} & $\thicksim$ \\
         \bottomrule
    \end{tabular}
    \label{tab:nice1_null-depth}
\end{table}

A comparison of the measured null depth and errors with the error budget is shown in \cref{tab:nice1_null-depth}.
During the measurement, closed-loop OPD control and shear control were active.
Shear control was likely not necessary, as even in open-loop, the measured error is far below the error budget.
The pointing metrology had not yet been implemented at the time of this measurement, but pointing was later found to be stable within \SI{10}{\micro\radian} RMS typically, even without an enclosure.
Intensity mismatch was not measured during the null, as we do not have access to the single-beam intensities after the spatial filter, and a measurement before the spatial filter would not include the varying coupling efficiency from pointing, shear, or wavefront errors.
Measuring the intensities of the beams sequentially after the null measurement was inconclusive because of drifts of the laser intensity during the recordings.
Since the measured null meets the requirement, we can however conclude that the intensity mismatch is also compliant with requirements.

\subsection{Stability}

There is a drift in null depth towards the end of the measurement, which was manually corrected using the delay line (not shown here).
The error is thus mostly in the OPD between the science beams, despite the metrology residuals not showing any error of this scale, indicating a drift between science-beam OPD and metrology-beam OPD.
We expect this to be a non-common-path error, for example caused by imperfect alignment between metrology and science beams, the different wavelengths between science and metrology, or the science beam traversing a path that is not visible to the metrology, such as the spatial filter.

Regardless of the source of these drifts in the null depth, we anticipate that moving towards cryogenic operation, with a much more stable temperature environment and with minimal lab-seeing, will increase the long-term stability of the system, moving from $\approx \SI{20}{\s}$ to our requirement of \SI{100}{\s}.
Any remaining drifts would then be corrected during an automated recalibration procedure, as described in \cref{ssec:nice1_duty-cycle-req}.

Regarding noise at higher frequencies in the power spectrum in \cref{fig:nice1_null-time-series}, peaks between \SIrange{10}{200}{\Hz} are identified with peaks measured by the OPD metrology, and are likely structural vibrations of the optics. Peaks at higher frequencies are due to intensity instabilities of the laser.
They are too small to have a significant impact on mean null depth, but a more stable laser will simplify intensity calibration.

\subsection{Spatial filtering}
\label{sec:discussion-spatial-filter}

The most likely source of the drifts in null depth is the spatial filter, as we have seen multiple problems associated with it.
There are two main issues we have identified with the single-mode fiber we used: leakge of higher order modes and sensitivity to external perturbations.

First, if the science beams are not well aligned before the spatial filter, we see a leakage of a tip/tilt mode through the cladding of the spatial filter, where moving the delay line slightly when in a null results in fringes moving laterally on the detector.
As a consequence, a null of \num{e-5} or less is then difficult to find, but can be regained after a re-alignment of the setup.
This indicates that the spatial filter has insufficient suppression of spatial modes for our purposes.
We emphasize here that the fiber core and cladding are likely single mode; rather, the interface between the cladding and coating of the fiber allow weak confinement of high-order modes at a level that still interferes with the null.
Such leakage has been investigated in previous nulling testbeds \cite{Ksendzov2007,Ksendzov2008}.
To quantify the allowed leakage and specify requirements on the single-mode fiber, consider that both beams just before the spatial filter have a static shear error $\xi$ and a static pointing error $\alpha$, and thus contain non-fundamental modes.
As derived in \cref{sec:nice1_derive-leakage}, this contributes a null depth leakage of
\begin{equation*}
    N_\text{leak}
    = \frac{1}{2} \frac{\eta_1}{\eta_0} \left( \frac{\xi^2}{w_0^2} + \frac{\alpha^2}{\theta_0^2} \right),
\end{equation*}
where $\eta_0$ is the throughput of the fundamental mode through the fiber, $\eta_1$ is the throughput of the first non-fundamental mode through the fiber, $w_0$ is the waist diameter of the beam, and $\theta_0$ the divergence, all measured before the coupling optics of the spatial filter.
As an example, assuming we want to maintain the capability to control intensity by \SI{20}{\percent} using shear and pointing offsets, and that we budget for a leakage null of \num{e-6}, the fiber has to achieve a power suppression ratio of $\eta_1/\eta_0 = \num{e-5}$.
A similar requirement of \num{e-6} for the suppression ratio was also imposed for spatial filters for the Darwin mission \cite{Flanagan_2006}.

Second, the spatial filter is extremely sensitive to external perturbations.
Even slightly touching the fiber while nulling results in a rise of intensity by a factor of $\approx 10$, indicating that some differential error between beams is introduced, possibly from higher-order modes getting stronger, or polarization errors being introduced.
Since the metrology beams do not traverse the spatial filters, any errors introduced therein cannot be not corrected by the control system.

To alleviate these issues, we are investigating alternative spatial filter solutions, such as cladding mode strippers \cite{Flatscher2006} or photonic crystal fibers (PCFs), also known as \enquote{endlessly single-mode fibers} \cite{Birks_1997,Ireland_2024}.
The former are components that attenuate high-order modes, usually through a custom, high-index fiber coating.
The latter shows promise in offering single-mode performance at longer wavelengths than the current \ce{InF3} fibers and can be much shorter, thus reducing the amount of non-common path between the metrology detectors and the science detector while exhibiting less sensitivity to environmental perturbations.
However, PCFs are at relatively low technological maturity, particularly in the mid-infrared wavelength band.
Further alternatives include hollow core fibers for the long wavelength bands \cite{Kriesel_2011}, the inclusion of a pinhole aperture at the fiber output \cite{Ksendzov2008}, and robust automated injection algorithms to reduce the initial strength of high-order modes at the coupling stage.

\subsection{Throughput}

While there are some caveats to the throughput measurement, which we will explain below, we argue to have demonstrated that the throughput requirement for NICE is achievable from an optical perspective for the shorter wavelength range, and that the remaining challenges for the Warm Bench are mostly on the opto-mechanical side and the alignment strategy.

First, we only have access to one of the two nulled outputs, since we are still developing a method to efficiently couple into two spatial filters simultaneously.
For a single output, we meet the requirement for the throughput (\SI{22}{\percent} measured vs \SI{17}{\percent} required), meaning optical interfaces and coatings are efficient enough.
Since the optical path of the second output would be nearly identical, the challenge is mostly opto-mechanical.

Second, we currently require a polarizer to hit a deep null, which would effectively halve our throughput if unpolarized light --- as would be expected from a planet --- was used.
We do not expect the throughput to differ for unpolarized light if the polarizer is removed, but achieving the null in both polarization modes simultaneously requires better alignment of the setup and matching of optical surfaces, or alternatively a method to control polarization errors.

Third, the measurement was performed at \SI{4.0}{\um}, representing the lower end of the LIFE science band.
The \ce{CaF2} beam-splitters are specified from \SIrange{2}{8}{\um}, and a broader wavelength range requires different materials or a splitting of the optical path into multiple bands.
The splitting ratio of the beam splitters, the reflectivity of the coatings, and the fiber losses will also vary as a function of wavelength; thus more effort is required to develop a broadband instrument.

\subsection{Comparison with other nulling testbeds}

\begin{table}
    \centering
    \footnotesize
    \caption[NICE compared with other nullers]{Comparison of first results on the NICE Warm Bench with a selection of other nulling testbeds.}
    \label{tab:nice-vs-other-nullers}
    \vspace{0.2 cm}
    \begin{tabular}{lScccS}
         \toprule
         Testbed & {Mean raw null depth} & Wavelength & Stability & Unpolarized? & {Throughput} \\
         \midrule
         NICE Warm Bench & \num{7.2e-6} & \SI{4.7}{\um} & \SI{25}{\s} & No & \SI{22 \pm 1}{\percent} \\[0.2 cm]
         PDT\cite{Martin_2012} & \num{3.8e-7} & \SI{10.6}{\um} & \SI{100}{\s} & No & {---} \\
         ANT\cite{Peters_2008} & \num{1.2e-5} & \SIrange{8}{12}{\um} & \SI{6}{\hour} & Yes & {---} \\[0.2 cm]
         SYNAPSE\cite{Gabor_2008_Tests} & \num{1.0e-5} & \SI{3.39}{\um} & \SI{10}{\minute} & No & {---} \\
         SYNAPSE\cite{Gabor_2008_Tests} & \num{1.5e-4} & \SIrange{2}{2.5}{\um} & \SI{10}{\minute} & Yes & {---} \\[0.2 cm]
         PERSEE\cite{LeDuigou_2017} & \num{8.8e-6} & \SIrange{1.65}{2.45}{\um} & \SI{100}{\s} & Yes & {---} \\[0.2 cm]
         MAII\cite{Weber_2004} & \num{1.3e-5} & \SI{1.55}{\um} & \SI{10}{\s} & No & {---} \\
         \bottomrule
    \end{tabular}
\end{table}

We compare the first results from the NICE Warm Bench with previous nulling testbeds in \cref{tab:nice-vs-other-nullers}.
NICE achieves raw mean nulls of $\mean{N} = \num{7.2e-6}$, which is similar to or slightly deeper than most (see also \cref{fig:nice1_history}).
This is expected, since the requirement on the raw null depth to detect and characterize terrestrial planets has remained relatively stable around \num{e-5} during these programs.
No existing nuller covers the LIFE science band of \SIrange{4}{18.5}{\um}, with ANT being the closest at \SIrange{8}{12}{\um}.
The stability of NICE of \SI{25}{\s} is on the shorter end of comparable setups, limited likely by the spatial filter, as discussed in \cref{sec:discussion-spatial-filter}, and has to be increased to at least \SI{100}{\s} to be compliant with requirements.
The high stability achieved by the PDT, the ANT, and SYNAPSE suggest that this should be feasible.
None of the major previous nulling testbeds have published throughput or sensitivity measurements, nor did they impose requirements on these parameters, so no direct quantitative comparison is possible.

\section{Outlook}
\label{sec:nice1_outlook}

With the Warm Bench having demonstrated sufficient throughput and null depth under simplified conditions --- a single output, a single polarization mode, and monochromatic light --- there are now multiple developments required in parallel to verify the LIFE beam combiner, which we will detail in this section in roughly chronological order. We summarize them, along with further technology developments, in \cref{tab:nice_tech_development}.

\begin{table}
    \centering
    \footnotesize
    \caption[Technology developments for NICE]{A summary of the technology developments related to NICE that are needed to successfully implement the LIFE mission's beam combiner.}
    \vspace{0.2 cm}
    \label{tab:nice_tech_development}
    \begin{tabular}{p{0.25\linewidth} p{0.7\linewidth}}
        \toprule
        \textbf{Technology} & \textbf{Description and Purpose} \\ \midrule
        Broadband nulling of unpolarized light & A beam combiner with low differential chromatic and polarization errors must be demonstrated. NICE currently uses a symmetric beam combination scheme, which results in two destructive outputs that must be corrected independently. \\ \midrule
        Broadband AR coatings & While transmissive substrates (e.g. \ce{KBr}, \ce{ZnSe}) are readily available, broadband anti-reflective coatings for beam splitters, dispersing prisms, and dichroics have to be developed and tested. \\ \midrule
        Optomechanics & To maintain a symmetric setup capable of deep nulls, the mounting of the optical components must be designed to remain stable while cooling down. Cryo-compatible actuators for compensating devices, such as linear, rotating, and tip/tilt stages, must be characterized.\\ \midrule
        Cryocoolers & A cryocooler chain must be coupled with low vibrations ($\lessapprox \SI{1}{\nm}$ OPD) to cool the combiner to $\SI{15}{\kelvin}$, or a potential superconducting detector to $< \SI{1}{\kelvin}$. \\ \midrule
        Cryogenic DM (optional) & For an adaptive nuller, cryo-compatible deformable mirrors with $\approx \SI{1}{\um}$ stroke and sub-$\si{\nm}$-stability must be tested and characterized. \\ \midrule
        Spatial filtering & Spatial filters that exhibit single-mode behavior over a large (or ideally all) of the LIFE bandpass must be developed. Requirements are efficient injection and propagation ($\approx \SI{60}{\percent}$ combined), leakgae-mode suppression better than $\approx \num{e5}$, and operation at \SI{15}{\kelvin}. \\ \midrule
        PIAA & Aspherical optics to anodize the incoming top-hat beam from the collector into a Gaussian beam for efficient coupling into the spatial filter. \\ \midrule
        Mid-infrared photonics (optional) & An integrated photonic beam-combiner with high throughput ($\approx \SI{50}{\percent}$) could replace the bulk-optics beam combiner and cross-combiner in NICE, and also serve as the spatial filter. Synergistic with an adaptive nuller to correct chromatic and polarization errors in the device for deep broadband nulls.\\ \midrule
        Fringe tracker & Implementation of an unambiguous interferometric phase measurement scheme to provide sub-\si{\nm} OPD estimates at $> \SI{1}{\hertz}$ bandwidth from unused light shorter than LIFE's science band.\\ \midrule
        Phase chopper and cross-combiner & For a four-beam double Bracewell combiner, a $\pm\pi/2$ phase chopping mechanism with $\approx \SI{1}{\minute}$ chopping period must be designed and tested. Such a device should not lead to disturbances that affect planet signal extraction. \\ \midrule
        Detectors & One or more detector technologies must be identified and procured that cover the LIFE bandpass with low dark current ($< 1 \, \text{e}^-/\text{s}$  and high ($> \SI{60}{\percent}$) quantum efficiency. This may include superconducting detectors that require $< \SI{1}{\kelvin}$ cooling. \\ \bottomrule
    \end{tabular}
\end{table}

\subsection{Broadband nulling of unpolarized light on the Warm Bench}
The required null has to be achieved with unpolarized light, a broadband source or multiple monochromatic sources, and in both outputs of the beam combiner, while retaining a throughput of more than \SI{34}{\percent}.

Sufficiently deep nulls with broadband unpolarized light have been achieved before, for example on the ANT \cite{Peters_2008} with an Adaptive Nuller, and preliminary internal raytracing studies indicate that this should be achievable on the Warm Bench without an adaptive nuller for restricted wavelength ranges with $\approx$ \SI{1}{\um} spectral bandwidth.
For broader bandwidths, options consist of compensation mechanisms or splitting into multiple wavelength bands (\cref{ssec:compensation}.)

While we expect null depth and spectral bandwidth to be similar to results from previous testbeds, we have more stringent requirements on throughput, and need to build a setup compatible with cryogenic operation.
At that point, the Warm Bench will have concluded its primary objective, but will remain in use for prototyping.

\subsection{Preparatory cryogenic studies}
Transitioning to a cryogenic setup is required to reach the sensitivity for a source of realistic planetary flux-levels.
Difficulties include alignment (cooling down may disturb the optical path), materials (matching CTEs, kinematic mounting) and vibrations (cryo-cooler, vacuum pumps), so early experimental verification is essential.
We are currently assembling a smaller test-cryostat with an internal volume of $\SI{30}{\cm} \times \SI{20}{\cm} \times \SI{20}{\cm}$ to characterize optics at \SIrange{10}{20}{\kelvin} on the level of single components and small assemblies.
The test cryostat is coupled with low- and higher-order wavefront sensors to characterize how optical assemblies, such as a mirror in a mount, deform while cooling down, which will inform the opto-mechanical strategy for the full NICE cryostat in the future.
We will also use it to characterize how vibrations from the cryocooler and vacuum pump are transmitted to the cold optics.

\subsection{Alignment strategy and compensation mechanisms}
\label{ssec:compensation}
Going from moderately broadband light to the full LIFE science band of \SIrange{4}{18.5}{\um} will require compensation of chromatic and polarization errors, caused by imperfect alignment or mismatches in coatings.
These errors are difficult to predict, so experimental verification is again paramount.
There are two approaches to compensating chromatic and polarization errors: active correction with an adaptive nuller, or lower-order independent compensation mechanisms.

The baseline approach is to achieve good static alignment before cooling down, and to compensate small residual errors with independent mechanisms, such as moving a compensator plate laterally for first-order control of chromatic errors, rotating the roof prism in the periscope to rotate polarization, and controlling ellipticity errors with a slight intentional misalignment in the tip/tilt stages.
Intensity as a function of wavelength can be controlled by introducing an intentional pointing error, which changes coupling efficiency into the spatial filter chromatically.
While these correction mechanisms are potentially simple to implement, they pose strict requirements on all optical tolerances, as they can only compensate for small low-order errors.

If these mechanisms should prove insufficient, the light can be split into polarization modes using a Wollaston prism, or into multiple wavelength bands using dichroic beam splitters, and multiple copies of the control section and the beam combiner can be used to control each subband individually.

A further possible extension of this method is the Adaptive Nuller \cite{Lay_2003}, which after splitting the light into polarization modes and wavelength bands, uses a deformable mirror (DM) to control OPD and intensity of each mode separately.
It is expected to loosen tolerances on coatings and alignment by orders of magnitude, and has the option to simplify and even make redundant many components in the optical path, such as removing the need for a second nulled output in the beam combiner.
Such a device has been successfully demonstrated with high contrast and stability \cite{Peters_2008} from \SIrange{8}{12}{\um}, but not at the high sensitivity and cryogenic temperatures required for NICE.

Promising cryogenic deformable mirrors based on Micro Electrical-Mechanical Systems (MEMS) have been demonstrated before \cite{Takahashi_2017,Enya_2009}, with 32 x 32 actuators and \SI{1}{\um} stroke at \SI{5}{\kelvin}.
Some open questions remain, such as their stability and achievable resolution under cryogenic conditions, whether the stroke is sufficient, how pixel defects can be dealt with, and how the wiring and the warm electronics close to the cold optics can be implemented.

\subsection{Spatial filtering and injection optics}

An efficient spatial filter is necessary to loosen requirements on wavefront quality.
While these devices exist off-the-shelf separately as narrowband mid-infrared devices and broadband visible-light devices\cite{Birks_1997}, an efficient spatial filter that works from \SIrange{4}{18.5}{\um} in the mid-infrared has not yet been experimentally demonstrated in a realistic environment.
Some single-mode fibers developed for Darwin have shown good transmission over a broad wavelength range, such as high-tellurium chalcogenide fibers with a measured attenuation of \SI{0.2}{\dB\per\cm} at \SI{10.6}{\um}, and Te-As-Se fibers with a measured transmission of \SI{0.1}{\dB\per\cm} at \SI{10.6}{\um} \cite{Cheng_2009}.
The former has also been tested at \SI{77}{\kelvin} from \SIrange{10}{15}{\um}, and transmission was found to improve across the full measurement band.
A \SI{20}{\cm} chalcogenide fiber with a mode suppression of $\eta_1/\eta_0 \approx \num{e-4}$ and \SI{0.08}{\dB\per\cm} loss at \SI{10.6}{\um} was characterized as part of the TPFI efforts \cite{Ksendzov2007}, which is the closest candidate to what is required for NICE for this wavelength range.
For NICE and LIFE, we consider splitting the science band into multiple sub-bands, with separate fibers optimized for each band.

Tightly coupled with the spatial filter itself are the injection optics, which in the case of LIFE have to efficiently couple a top-hat beam into a waveguide with a nearly Gaussian fundamental mode.
Techniques such as phase-induced amplitude apodization (PIAA) \cite{Guyon_2003,Jovanovic_2017}, which can reshape the beam into a near-Gaussian profile, are promising not only to increase coupling, but also extend the broadband response via a wavelength dependent beam diameter. 
Both efficient broadband spatial filters and PIAA injection optics are currently under investigation.\cite{Ireland_2024}

\subsection{Photonic integrated devices}
Photonic integrated devices, essentially an optical setup on a microchip, may offer great potential for miniaturization, increased stability, and repeatable modular manufacturing.
They are commonly used in modern ground-based nullers (NOTT\cite{Garreau_2024,Sanny_2026}, GLINT\cite{Norris_2019,Martinod_2021}) for these reasons, but they are almost entirely unexplored beyond \SI{4}{\um} because of a lack of transparent substrates and mature processes, with recent developments pushing towards the longer end of LIFE's bandpass\cite{MontesinosBallester_2024}.

\subsection{Missing subsystems}
Currently, we only simulate an on-axis star, which we aim to extinguish through nulling.
A realistic demonstration of the instrument should measure the throughput of an off-axis planet while nulling the star, which requires a planet simulator.
Furthermore, LIFE cannot rely on internal metrology systems to measure the path from the star to the collector mirrors, so a fringe tracker has to be demonstrated, using a source at $\approx \SI{2}{\um}$.
Finally, LIFE is a Double-Bracewell nuller, while NICE is currently a Single-Bracewell design.
A copy of the setup and a final cross-combiner stage are needed to verify the full four-beam architecture.

\section{Conclusion}
\label{sec:nice1_conclusion}

We are building the Nulling Interferometry Cryogenic Experiment (NICE), a technology demonstrator and development testbed for the beam combiner of the LIFE space mission.
NICE is designed as the first cryogenic nuller that can achieve the sensitivity and contrast required to characterize terrestrial exoplanets in the mid-infrared.
While the requirements for NICE are not yet fully derived from the LIFE science case and the LIFE reference design in a systematic manner, we argue that the requirements we have set for NICE are likely more ambitious than necessary to ensure compliance in the future.
NICE requires a $\approx \num{e-5}$ null depth from \SIrange{4}{18.5}{\um} while maintaining a high throughput of \SI{34}{\percent} in a cryogenic environment.
At ambient conditions, using a narrowband \SI{4.7}{\um} source and a polarizer at the output, we have achieved a mean raw null depth of \SI{7e-6} over \SI{25}{s}, deeper than any previously published result in the M-band, as well as a throughput of \SI{22}{\percent} from one of the two nulled outputs of the beam combiner.
Both null depth and throughput are compliant with requirements.
We are extending the Warm Bench for more broadband nulls with unpolarized light in the near future.

While the principle of the measurement has thus been demonstrated in a simplified environment, and the strategy for a sensitive cryogenic instrument has been laid out, we also emphasize some technological gaps that lie ahead of us.
Some of the subsystems required for NICE, such as the spatial filter and the compensation mechanisms required for nulling over two octaves of spectral bandpass, while conceptually defined, are still experimentally unverified.
Further studies are required before NICE can transition to a cryogenic setup.

\appendix

\section*{Appendix}

\subsection*{Use of AI tools}
Github Copilot was used as an auto-complete tool while writing the code for data analysis and numerical evaluation of the error budget.
No other AI tools were used in the preparation of this work.

\subsection*{Disclosures}
The authors declare that there are no financial interests, commercial affiliations, or other potential conflicts of interest that could have influenced the objectivity of this research or the writing of this paper.

\subsection*{Code, Data, and Materials Availability} 
The data presented in this article, as well as the source code to analyze the data and produce the tables and figures in this article, are publicly available on Github at \url{https://github.com/thomabir/NICE-Paper-I}.

\subsection*{Acknowledgements}
Part of this work has been carried out within the framework of the National Centre of Competence in Research PlanetS supported by the Swiss National Science Foundation under grants \texttt{51NF40\_182901} and \texttt{51NF40\_205606}.
NICE is financially supported by the Swiss Prodex program directed by the Swiss Space Office. This project is supported by Rudolf Bär via the ETH Zurich Foundation.

\section{Derivation of the error budget}
\label{sec:nice1_derive-error-budget}

Here, we derive how the mean null depth during an observation can broken down into individual error terms.
The goal is not to tolerance the optical setup in detail, but to derive the higher-level error terms that can later be used for optical tolerancing.
Much of the derivation follows the work of Serabyn\cite{Serabyn_2000}, with some modifications, such as not assuming a lossless 50/50 beam-splitter, and also breaking down the intensity mismatch into pointing and shear errors.
We present the derivation here mostly for the sake of consistency and easy reference.

\subsection{Null depth for a single polarization mode}

For a single wavelength, polarization mode, and spatial mode, the only quantities that affect the null depth are the intensities $I_1$, $I_2$, and the phases $\phi_1$, $\phi_2$ of the two beams.
The electric field amplitudes at the detector for two arbitrary beams are then
\begin{align*}
    E_1 = \sqrt{I_1} \exp(i \phi_1), \qquad
    E_2 = \sqrt{I_2} \exp(i \phi_2).
\end{align*}
We define the maximum constructive intensity $I_+ = I_1 + I_2 + 2 \sqrt{I_1 I_2}$, the phase error $\delta\phi = \abs{\phi_1 - \phi_2}$, and the intensity mismatch $\delta I = \abs{I_1 - I_2} / (I_1 + I_2)$.
The instantaneous nulled intensity at the destructive output is then $I_- = \abs{E_1 - E_2}^2$, and the null depth is
\begin{align*}
    N = \frac{I_-}{I_+}
    = \frac{1-\sqrt{1-\delta I^2} \cos(\delta\phi)}{1+\sqrt{1-\delta I^2}}
    \approx \frac{1}{4} (\delta\phi^2 + \delta I^2),
\end{align*}
where the final expression is the second-order Taylor approximation for $\delta I, \delta\phi$ around zero.

\subsection{Null depth for both polarization modes}

Consider the case of two orthogonal polarization modes, such as $s$ and $p$, still for a single wavelength and spatial mode,
\begin{equation*}
    N_s
    \approx \frac{1}{4}(\delta\phi_s^2 + \delta I_s^2), \qquad
    N_p
    \approx \frac{1}{4}(\delta\phi_p^2 + \delta I_p^2),
\end{equation*}
where both modes may have independent errors in phase and intensity.
The null depth when both modes are combined on the same detector is the mean of the two null depths,
\begin{equation}
    \label{eq:null-inst-jones}
    N = \frac{N_s + N_p}{2} =\frac{1}{4} \Big( \delta\phi^2 + \frac{1}{4}\delta\phi_{sp}^2 + \delta I^2 + \frac{1}{4} \delta I_{sp}^2 \Big),
\end{equation}
where
\begin{align*}
    \delta \phi &= \frac{\delta \phi_s + \delta \phi_p}{2}, && \text{Average phase error in s and p mode},\\
    \delta \phi_{sp} &= \abs{\delta \phi_s - \delta \phi_p}, && \text{Difference in phase mismatch between s and p mode},\\
    \delta I &= \frac{\delta I_s + \delta I_p}{2}, && \text{Average intensity mismatch in s and p mode},\\
    \delta I_{sp} &= \abs{\delta I_s - \delta I_p}, && \text{Difference in intensity mismatch between s and p mode}.\\
\end{align*}

\subsection{Dynamic top-level error budget}

In a real instrument, errors will vary over time.
To produce a first-order estimate of the resulting mean null depth, we assume that all errors are independent white Gaussian noise over time, and that errors are small, so the lowest non-zero terms in the Taylor series are used.
We further assume that all polarization errors are static in time.
This results in
\begin{equation}
    \mean{N}(\lambda)
    = \frac{1}{4} \Big( \mean{\delta\phi}^2 + \sigma_{\delta\phi}^2 + \frac{1}{4}\mean{\delta\phi_\text{sp}}^2 + \mean{\delta I}^2 + \sigma_{\delta I}^2 + \frac{1}{4} \mean{\delta I_\text{sp}}^2 \Big),
\end{equation}
where angled brackets $\langle \dots \rangle$ denote the mean over time, and $\sigma$ denotes the sample standard deviation over time.
This equation is used to derive the equal-weight top-level error budget in \cref{tab:nice1_error-budget}.

\subsection{Influence of pointing and shear errors}

Any change in the shear and pointing of the science beams leads to a change in coupling efficiency into the spatial filter, which will affect the null depth.
To derive the impact of these errors, we first derive how a static shear or pointing offset affects throughput and null depth, and then derive the dynamic effect.
We assume Gaussian beams and waveguides with Gaussian modes.

We use the Fresnel approximation to analytically model the action of focusing with a lens.
The field $E_1$ at the focal plane of the lens is then related to the field $E_0$ just before the lens by\cite[p.~103]{Goodman_1996}
\begin{equation}
    \label{eq:nice1_focus-lens}
	E_1(x,y) =
	- \frac{i}{f \lambda}
	\exp\left[
	\frac{i \pi}{f \lambda}(x^2 + y^2)
	\right]
	 \iint_{\R^2}
	E_0(x',y') \exp \left[-
		\frac{2 \pi i}{f \lambda}
		( x x' + y y' )
	\right]
	\, \mathrm{d}x' \, \mathrm{d}y',
\end{equation}
where $\lambda$ is the wavelength, and $f$ the focal length of the lens.

Coupling into a single mode fiber is modeled by the overlap integral of the electric field $E_1$ at the entrance of the fiber with the Gaussian mode $\psi$ of the fiber.
The fiber is characterized by its mode field radius $w_f$, which is the waist radius of the Gaussian beam that propagates in the fiber.
The normalized fiber mode is then
\begin{align*}
    \psi(x,y) = \sqrt{\frac{2}{\pi w_f^2}} \exp\left( -\frac{x^2 + y^2}{w_f^2} \right),
\end{align*}
where \emph{normalized} means that $\iint_{\R^2} \psi(x,y)^2 \, \mathrm{d}x \, \mathrm{d}y = 1$.
Before defining the coupling efficiency, we introduce the overlap integral
\begin{align*}
    \iprod{E_1}{\psi} = \iint_{\R^2} E_1(x,y) \, \psi^*(x,y) \, \mathrm{d}x \, \mathrm{d}y,
\end{align*}
where $\psi^*$ denotes the complex conjugate of $\psi$.
The coupling efficiency $\eta$ is then defined as the power ratio of input beam and the output beam of the fiber,\cite[p.~21.8]{Bass_2010}
\begin{align}
    \label{eq:nice1_fiber-coupling-efficiency}
    \eta = \frac{
        \abs{\iprod{E_1}{\psi}}^2
    }{
        \iprod{E_1}{E_1^*}
        \iprod{\psi}{\psi^*}
    }.
\end{align}
We neglect here effects such as Fresnel reflection at the fiber facet and propagation losses, since we are only interested in how a change in pointing or shear changes the throughput, not in absolute throughput.

A Gaussian beam propagating along the $z$-axis has at its waist, i.e. its narrowest point, an electric field
\begin{align}
\label{eq:nice1_gauss-beam}
	E(x,y) = E_0 \exp \left(-\frac{x^2+y^2}{w_0^2} \right),
\end{align}
where $x^2+y^2 = r^2$ is the squared distance from the optical axis,
$E_0$ is the electric field at the origin, and
$w_0$ is the waist radius, which is the radius where the amplitude falls off to $1/e$.

A Gaussian beam that is tilted by a small angle $\alpha$ along the $x$-axis is obtained by adding a linear phase shift $k x \alpha$ to \cref{eq:nice1_gauss-beam},
\begin{align}
    \label{eq:nice1_gauss-tilt}
	E_0(x,y) = E_0 \exp\left(
		- \frac{x^2+y^2}{w_0^2}
		+ i k x \alpha
	\right).
\end{align}
We set $E_0 = 1$ from now, since the amplitude would anyway cancel out in the throughput calculation.
To find how a tilted Gaussian beam propagates to the focal plane, we plug \cref{eq:nice1_gauss-tilt} for $E_0$ into \cref{eq:nice1_focus-lens} and evaluate the integral, which yields
\begin{align}
    \label{eq:nice1_focused-tilted}
    E_1(x,y) = - \frac{i \pi  w_0^2}{f \lambda }
    \exp \left(-\frac{1}{4} \alpha ^2 k^2 + \frac{\pi  \alpha  k w_0^2 x}{f \lambda } - \frac{\pi^2 w_0^2 (x^2+y^2)}{f^2 \lambda ^2}\right),
\end{align}
where a term $i \pi r^2 / (f \lambda)$ in the exponent was dropped as it is small compared to the other terms.
To calculate the coupling efficiency, we plug \cref{eq:nice1_focused-tilted} into the \cref{eq:nice1_fiber-coupling-efficiency}, and evaluate:
\begin{align}
    \label{eq:nice1_eta-tilted}
    \eta_\text{pointing} = \exp\left( -\frac{\alpha ^2}{\theta_0 ^2} \right)
    = 1 - \frac{\alpha ^2}{\theta_0^2} + \bigO(\alpha^4),
\end{align}
where $\theta_0 = \lambda / (\pi w_0)$ is the divergence of the Gaussian beam.
We used here that $f = \pi w_0 w_f / \lambda$ is the focal length required to focus the initial beam with waist $w_0$ onto the fiber profile with waist $w_f$, which leads to all occurrences of $w_f$ canceling out.

If the beam has a lateral offset $\xi$ with respect to the optical axis along the $x$-direction, then the initial beam profile is
\begin{align*}
	E_0(x,y) = E_0 \exp\left[
		- \frac{(\xi-x)^2 + y^2}{w_0^2}
	\right].
\end{align*}
We follow the same procedure as for the tilted beam to arrive at the throughput
\begin{align}
	\eta_\text{shear} = \exp\left(-\frac{\xi^2}{w_0^2}\right)
	= 1 -\frac{\xi^2}{w_0^2} + \bigO(\xi^4).
\end{align}

The corresponding intensity mismatch that would result from an error of this kind, assuming a single beam is affected, is
\begin{equation}
    \delta I = \frac{1-\eta}{1+\eta}
    = \tanh\left( \frac{\chi^2}{2} \right)
    = \frac{\chi^2}{2} + \bigO(\chi^4),
\end{equation}
where the functional form $\eta = \exp(-\chi^2)$ is assumed.
This yields
\begin{equation}
    \delta I_\text{shear} \approx \frac{\xi^2}{2 w_0^2}, \qquad
    \delta I_\text{pointing} \approx \frac{\alpha^2}{2 \theta_0^2}.
\end{equation}

To translate from static errors into dynamic errors, we use higher-order Gaussian error propagation:
\begin{align}
    \sigma_{\delta I}^2
    = \sigma_\chi^2 \left( \dv{\delta I}{\chi} \right)^2
    + \frac{1}{2} \sigma_\chi^4 \left( \dv[2]{\delta I}{\chi} \right)^4
    + \bigO(\sigma_\chi^6)
    \approx \frac{\sigma_\chi^4}{2},
\end{align}
where we used that the mean of $\chi$ is zero (no DC error).
For a combined error in shear and pointing, we plug in for $\chi$ to find
\begin{equation}
    \sigma_{\delta I}^2 = \frac{1}{2}\left( \frac{\sigma_\xi}{w_0} \right)^4 + \frac{1}{2}\left( \frac{\sigma_\alpha}{\theta_0} \right)^4,
\end{equation}
which is used to derive the contribution of shear and pointing to the error budget in \cref{tab:nice1_error-budget}.

\section{Derivation of null leakage from an imperfect spatial filter}
\label{sec:nice1_derive-leakage}

We want to derive how a Gaussian beam with a small static error $\xi$ in shear and a small static error $\alpha$ in pointing  propagates through an imperfect spatial filter, and how the null depth is affected by this.
We first derive how the field of the perturbed beam can be decomposed into modes, and then propagate these modes through the spatial filter to find how they degrade the null depth.

\subsection{Modal decomposition}

We define the fundamental Gaussian mode $\psi_{00}$ and the tip/tilt mode $\psi_{10}$ as
\begin{align*}
    \psi_{00}(x,y) &= \sqrt{\frac{2}{\pi w_0^2}} \exp\bigg(\!-\frac{x^2 + y^2}{w_0^2} \bigg),\\
    \psi_{10}(x,y) &= \frac{2 x}{w_0} \psi_{00}(x,y),
\end{align*}
where $w_0$ is the waist radius of the Gaussian beam.
These are the lowest-order Hermite-Gaussian modes, and they are normalized in power, such that
\begin{equation*}
    \int_{\R^2} \psi_{00}^2 \dd{x} \dd{y}
    = \int_{\R^2} \psi_{10}^2 \dd{x} \dd{y}
    = 1.
\end{equation*}
An unperturbed Gaussian beam in front of the spatial filter optics has a field $E(x,y) = E_0 \psi_{00}(x,y)$, and a total power of $\int_{\R^2} E^2 = E_0^2 = I_0$.

If a beam is Gaussian but with a small pointing error $\alpha$ in the $x$-direction, its field is
\begin{align*}
    E_\alpha(x,y)
    &= E_0 \, \psi_{00}(x,y) \, e^{i k \alpha x}\\
    &\approx E_0 \, \psi_{00}(x,y) \, (1 + i k \alpha x)\\
    &= E_0 \! \left[ \psi_{00}(x,y) + i \frac{\alpha}{\theta_0} \psi_{10}(x,y) \right],
\end{align*}
where $\theta_0 = \lambda / (\pi w_0)$ is the divergence of a Gaussian beam.

If the beam instead has a small shear error $\xi$, then the field is
\begin{align*}
    E_\xi(x,y)
    &= E_0 \, \psi_{00} (x - \xi, y) \\
    &\approx E_0 \left[ \psi_{00}(x,y) - \xi \pdv{\psi_{00}}{x} \right] \\
    &= E_0 \left[ \psi_{00}(x,y) + \frac{\xi}{w_0} \psi_{10}(x,y) \right].
\end{align*}

If both errors are present, we can read off the coefficient $c_{01}$ of the first mode as
\begin{equation*}
    c_{01}
    = \frac{\xi}{w_0} + i \frac{\alpha}{\theta_0},
\end{equation*}
and the power in this mode is thus
\begin{equation*}
    I_{01}
    = I_0 \left( \frac{\xi^2}{w_0^2} + \frac{\alpha^2}{\theta_0^2} \right).
\end{equation*}
The optics for coupling into or out of the spatial filter do not change the power of this mode, since focusing the beam with a lens merely swaps pointing and shear.

\subsection{Effect on null depth}

Assume both beams coupling into the spatial filter have equal intensity $I_0$ (a good approximation when close to a null), that they both have worst-case errors $\xi$ and $\alpha$, and that the fiber has a throughput of $\eta_0$ for mode $\psi_{00}$ and a throughput of $\eta_1$ for mode $\psi_{10}$.
We further assume as a conservative estimate that the two modes are incoherent after having propagated to the fiber, and thus $\psi_{10}$ fully contributes to the leakage.
The power $I_\text{leak}$ of the $\psi_{10}$ mode after the fiber can now be modeled as
\begin{align*}
    I_\text{leak} = 2 \eta_1 I_0 \left( \frac{\xi^2}{w_0^2} + \frac{\alpha^2}{\theta_0^2} \right),
\end{align*}
and the resulting leakage null depth is
\begin{align*}
    N_\text{leak}
    = \frac{I_\text{leak}}{I_1 + I_2 + 2 \sqrt{I_1 + I_2}}
    = \frac{1}{2} \frac{\eta_1}{\eta_0} \left( \frac{\xi^2}{w_0^2} + \frac{\alpha^2}{\theta_0^2} \right),
\end{align*}
where we used the definition of null depth from \cref{eq:nice1_null-depth-definition}, and that the total powers of the single beams after the fiber are $I_1 \approx I_2 \approx I_0 \eta_0$.
The ratio $\eta_1/\eta_0$ is the ratio by how much the fiber suppresses higher-order mode light compared to fundamental-mode light.


\bibliography{bibliography}   
\bibliographystyle{spiejour}   

\listoffigures
\listoftables

\end{spacing}
\end{document}